\documentclass[reprint,superscriptaddress, amsmath,amssymb, aps, prb,floatfix]{revtex4-1}

\usepackage{graphicx}
\usepackage{dcolumn}
\usepackage{bm}
\usepackage{hyperref}
\usepackage{xcolor}
\usepackage{lipsum}
\usepackage{transparent}

\begin{document}
\bibliographystyle{apsrev4-1}
\title{Short-range charge fluctuations in the two-dimensional Hubbard model}

\author{Xinyang Dong}
\affiliation{Department of Physics, University of Michigan, Ann Arbor, MI 48109, USA}
\author{Emanuel Gull}%
\affiliation{Department of Physics, University of Michigan, Ann Arbor, MI 48109, USA}

\begin{abstract}
We investigate charge fluctuations in the two-dimensional Hubbard model as a function of doping, interaction strength, next-nearest-neighbor hopping, and temperature within the eight-site dynamical cluster approximation. In the regime of intermediate interaction strengths, we find that $d$ density wave fluctuations, which had previously been postulated using analytical arguments, are present and strong but cannot be interpreted as the cause of the pseudogap in the model, due to their evolution with doping and interaction strength. For all parameters away from half filling, the charge fluctuations investigated, including $d$ density wave fluctuations, are weaker than $d$-wave superconducting fluctuations.
\end{abstract}

\date{\today}
\maketitle
\section{Introduction}

The presence of large competing fluctuations of several kinds is one of the defining aspects of correlated electron systems. These fluctuations may then condense into phases that exhibit remarkable properties, including unusually high superconducting transition temperatures and interesting magnetism.

The cuprate superconductors are a paradigmatic example for such a competition.  Antiferromagnetic fluctuations are strongest in the `undoped' parent compounds but present over a large part of phase space. Superconducting fluctuations lead to a superconducting dome for dopings smaller than $\sim 20\%$. Charge phenomena  \cite{Fujita04_stripe, Kohsaka07_stripe, Parker10, Lawler10_stripe, Hashimoto10_stripe, Daou10_stripe, Abbamonte12_stripe, Tranquada12_stripe, Abeykoon13_stripe, Jacobsen15_stripe, Merritt19_stripe}, such as the famous stripes at 1/8th doping \cite{Fujita04_stripe, Tranquada12_stripe} or the features observed in scanning-tunneling experiments \cite{Lawler10_stripe}, are present in several parts of phase space. Charge order with d-wave symmetry has also been found in RXS \cite{Comin2015}.

A minimal model that describes many of the salient features of these materials is the single-band Hubbard model \cite{Anderson87, Scalapino07}. While unable to describe excitations involving high-lying orbitals, the model reproduces much of the observed low-energy phenomenology, including a pseudogap (PG) \cite{Huscroft01,Civelli05,Stanescu06,Parker08,Liebsch09,Ferrero09,Ferrero09B,Sakai09,Sakai10,Lin10}, superconductivity \cite{Maier00, Civelli09, Civelli09B, Gull12_energy, Gull13_super, Gull13_raman, Gull14_pairing, Gull15_qp, Chen2015}, and the response functions of Raman spectroscopy \cite{Lin12,Gull13_raman}, optical conductivity \cite{Ioffe00,Toschi05,Millis05,Comanac08,Ferrero10,Lin10,Bergeron11}, nuclear magnetic resonance \cite{Macridin06,Chen17}, and neutron spectroscopy \cite{LeBlanc19_neutron}. It is therefore interesting to examine properties of the model in the context of cuprate physics. Due to the non-perturbative parameter regime relevant to the materials, reliable predictions have to resort to numerics, and a wide range of efficient numerical methods are able to describe the relevant parameter regime with consistent results \cite{LeBlanc15}.

Hubbard model calculations find spin, charge, and superconducting fluctuations. Spin fluctuations are well understood and mainly dominant near half filling \cite{Sherman2007, Hochkeppel08, Devereaux15, LeBlanc19_neutron}. Calculations also find that strong short-wavelength spin fluctuations are  primarily responsible for the formation of the pseudogap \cite{Gull09_8site,Gunnarsson15,Wu2018,Dong2019}, {\it i.e.}, the suppression of the density of states near the antinode but not near the node. Superconductivity is found unambiguously in the weak coupling regime \cite{Zanchi96,Honerkamp01,Deng15}, and strong indications from dynamical cluster calculations show that superconductivity does persist to larger couplings \cite{Maier05_dwave}. In contrast, results from some newer methods find that in the absence of a next-nearest-neighbor hopping, it is charge (rather than superconducting) order that dominates the ground state \cite{Zheng17_stripe,Qin19}. However, all orders are in very close competition.
The precision to which the energetics of these phases is known is much better than the uncertainty in the model parameters, indicating that phenomena beyond simple Hubbard model physics may well force the system to choose one order over the other. 

At finite temperature, charge fluctuations, in contrast to antiferromagnetic and superconducting fluctuations, are less well investigated for the model without additional nearest-neighbor interactions, whereas the ``extended'' model has been studied extensively in recent years \cite{Aichhorn04,Husemann12,Ayral13,vanLoon14,Huang14,vanLoon16,Kapcia17,Medvedeva17,Jiang18,Terletska18,Paki19,Pudleiner19,Schueler19}. This is despite the fact that theoretical approaches have proposed unusual charge phenomena, such as the $d$ density wave (DDW) order \cite{Nayak2000}, as candidates responsible for pseudogap physics \cite{Chakravarty2001, Tewari2001, Morr2002, Morr2003}. It is therefore interesting to investigate the extent to which charge fluctuations are present in the model, and the extent to which they correspond to the proposed $d$ density wave fluctuations, using numerical methods that generate these fluctuations dynamically from an underlying Hamiltonian. As we shall show below, the results are unexpected. First, we do find substantial $d$ density wave fluctuations. However, while we find no area in parameter space where those fluctuations are dominant, they are comparable in magnitude to superconducting fluctuations (DSC). Also, while fluctuations are large in the general area of the pseudogap, the behavior with doping, interaction strength, and next-nearest neighbor hopping is not consistent with $d$ density wave fluctuations as a mechanism for the suppression of the density of states.

The remainder of this paper is organized as follows. In Sec.~\ref{sect: method}, we describe the method used in this paper. In Sec.~\ref{sect: fluct}, we show the comparison between leading fluctuations in the 2D Hubbard model. In Sec.~\ref{sect: temp}, we show the temperature evolution of DDW and DSC. In Sec.~\ref{sect: ph}, we show the next-nearest-neighbor hopping evolution of DDW. Sections~\ref{sect: diss} and \ref{sect: conclu} discuss our results and present conclusions. 

\section{Method} \label{sect: method}
We investigate the two-dimensional Hubbard model on a square lattice with on-site interaction $U$ and chemical potential $\mu$,
\begin{equation}
   H=\sum_{k\sigma}(\varepsilon_k-\mu)c^\dagger_{k\sigma}c_{k\sigma}+U\sum_in_{i\uparrow}n_{i\downarrow}. \label{Eq:Hubbard}
\end{equation}
Here $i$ labels the lattice site, $k$ the momentum, $c^{(\dagger)}$ annihilation (creation) operators, and $n$ the density. $\varepsilon_k=-2t(\cos k_x + \cos k_y) -4t' \cos k_x \cos k_y$ is the dispersion with hopping $t$ and next-nearest-neighbor hopping $t'$. 

We define the single particle Green's function as 
$G_\sigma(k_1\tau_1,k_2\tau_2) $ $=$ $ \langle T_\tau (c_{k_1\sigma}^\dagger (\tau_1) c_{k_2\sigma}(\tau_2))\rangle$
and the two-particle Green's function as 
$G_{2, \sigma_1\sigma_2\sigma_3\sigma_4}(k_1\tau_1,...,k_4\tau_4) $ $=$ $ \langle T_\tau (c_{k_1\sigma_1}^\dagger (\tau_1) c_{k_2\sigma_2}(\tau_2) c_{k_3\sigma_3}^\dagger (\tau_3) c_{k_4\sigma_4}(\tau_4))\rangle$, , with conservation of momentum $k_1 + k_3 = k_2 + k_4$, and  $\tau_i$ denote imaginary time points.

The generalized susceptibility is defined as \cite{Roheringer12} $\chi_{\sigma \sigma'}(k_1\tau_1,k_2\tau_2,k_3\tau_3,k_4 0)$ $=$ $G_{2,\sigma\sigma\sigma'\sigma'}(k_1\tau_1,k_2\tau_2,k_3\tau_3,k_4 0)$ $-$ 
$G_{\sigma}(k_1\tau_1,k_2\tau_2)G_{\sigma'}(k_3\tau_3,k_4 0)$.
Its Fourier transform in the particle-hole notation is
\begin{align}
    &\chi_{ph, \sigma \sigma'}^{\omega\omega'\Omega}(k,k',q) = \int_{0}^{\beta} e^{-i\omega\tau_1}e^{i(\omega+\Omega) \tau_2}e^{-i(\omega'+\Omega) \tau_3} \nonumber\\&\times \chi_{\sigma\sigma'}(k\tau_1,(k+q)\tau_2,(k'+q)\tau_3,k' 0) d\tau_1 d\tau_2 d\tau_3. \label{eq:chiph}
\end{align}
The ``density channel" susceptibility is then defined as
\begin{align}
    \chi_{d}^{\omega\omega'\Omega}(k, k', q)=\chi_{ph, \uparrow\uparrow}^{\omega\omega'\Omega}(k, k', q) +\chi_{ph, \uparrow\downarrow}^{\omega\omega'\Omega}(k, k', q),
\end{align}
and can be decomposed using the Bethe-Salpeter equation into
\begin{align}
    \chi_{d}^{\omega\omega'\Omega}&(k, k', q) = \chi_0^{\omega\omega'\Omega}(k, k', q) - \frac{1}{\beta^2 N^2}\chi_0^{\omega\omega_1\Omega}(k, k_1, q)\nonumber \\
    &\times \Gamma_{d}^{\omega_1\omega_2\Omega}(k_1, k_2, q)\chi_d^{\omega_2\omega'\Omega}(k_2, k', q),
    \label{eq:chid}
\end{align}
with $\chi_0^{\omega\omega'\Omega}(k, k', q) = -\beta N G_\sigma (i\omega, k)G_\sigma (i\omega+i\Omega, k+q)\delta_{\omega\omega'}\delta_{k k'}$ the bare susceptibility and $\Gamma_{d}^{\omega_1\omega_2\Omega}(k_1, k_2, q)$ the irreducible vertex in the density channel. $N$ denotes the number of $k$ points for the summation over $k_1,~k_2$.
%
%

Linear response theory relates $\chi_d$ to a generating field $\Lambda(k)$ as \cite{Baym1961}
\begin{equation}
\begin{aligned}
    -\frac{1}{N_0}\int_0^\beta d\tau' \sum_{k k'} \frac{\delta G_\sigma(k+q,0,k,0; \Lambda)}{\delta \Lambda(k', \tau')} g(k)g(k')\bigg \vert_{\Lambda=0} &
    \\= \frac{1}{\beta^2 N_0}\sum_{\omega\omega',k k'}\chi_{d}^{\omega\omega'0}(k, k', q)g(k)g(k')
    \label{eq:chisymm}
\end{aligned}
\end{equation}
with $N_0 = \sum_{k k'}|g(k)g(k')|$ a normalization and $g(k)$ a symmetry factor. The related order parameter is $D = i \sum_{k, \sigma} g(k) c_{k+q,\sigma}^\dagger c_{k,\sigma}$ \cite{Chakravarty2001, Macridin2004}.

Superconducting fluctuations are related to the susceptibility in the particle-particle channel which follows from an analogous derivation \cite{Chen2015}, 
\begin{equation}
    \begin{aligned}
    \frac{1}{N_0}\int_0^\beta d\tau' \sum_{k k'} \frac{\delta F(k,0; \eta)}{\delta \eta(k', \tau')} g(k)g(k')\bigg \vert_{\eta=0} &
    \\= \frac{1}{\beta^2 N_0}\sum_{\omega\omega',k k'}\chi_{pp \bar{\uparrow\downarrow}}^{\omega\omega'0}(k, k', q = 0)g(k)g(k')
    \label{eq:sc}
\end{aligned}
\end{equation}
with $F(k, \tau)=-\langle T_{\tau} c_{k, \uparrow}(\tau) c_{-k, \downarrow}(0) \rangle$ the anomalous Green's function, and $\eta(k)$ the generating field. The related order parameter is $P = \sum_{k} g(k) c_{k,\uparrow} c_{-k,\downarrow}$ \cite{Macridin2004}.

We define the right hand side quantity in both Eq.\ref{eq:chisymm} and Eq.~\ref{eq:sc} as $\chi_g$,
\begin{align}
    \chi_g = \frac{1}{\beta^2 N_0}\sum_{\omega\omega',k k'}\chi^{\omega\omega'0}(k, k', q)g(k)g(k').
    \label{eq:chig}
\end{align}
This is the central quantity investigated in this paper.

Phase transitions are indicated by a divergence of the susceptibility $\chi^{\omega,\omega',\Omega}$. Since we expect large values of $\chi^{\omega,\omega',\Omega}$ to be caused by vertex contributions, we define
\begin{align}
    V_g := &(\chi - \chi_0)_g = \frac{1}{\beta^2 N_0}\sum_{\omega\omega',k k'} g(k)g(k') \, \times \nonumber\\
    &(\chi^{\omega\omega'0}(k, k', q) - \chi_{0}^{\omega\omega'0}(k, k', q))
    \label{eq:vertsymm}
\end{align}
in analogy to $P_g$ in the superconducting case \cite{Chen17}.
This quantity highlights vertex contributions by subtracting band-structure and single-particle effects contained within the bare susceptibility.

We will show that charge fluctuations with momentum transfer $q = (\pi, \pi)$ and d-wave symmetry are large.
These fluctuations are known as $d$ density waves \cite{Nayak2000, Chakravarty2001}.
Ref.~\onlinecite{Nayak2000} defines density order in analogy with superconductivity, with a singlet order parameter in the form $\langle \psi_\alpha ^\dagger(k+Q) \psi_\beta(k) \rangle = \Phi_Q f(k) \delta_{\alpha,\beta}$ for symmetry factor $f(k)$.
Two of the possible orders are $f(k) = \sin k_x$ for $p_x$ symmetry and $f(k) = \cos k_x - \cos k_y$ for $d_{x^2-y^2}$ symmetry.
The d density wave, also called the ``staggered flux state", occurs at $Q = (\pi, \pi)$. The $p_x$ density wave (PDW), also called the ``bond order wave", happens at $Q = (\pi, \pi)$ and $(\pi, 0)$. For a real space representation of DDW, see Fig.~2 of Ref.~\onlinecite{Nayak2000} and Fig.~2 of Ref.~\onlinecite{Chakravarty2001}.

The exact solution of the Hamiltonian Eq.~\ref{Eq:Hubbard} is unknown. Here we use the dynamical cluster approximation (DCA) on a cluster with $N_c = 8$ sites, which approximates the self-energy by $N_c$ “coarse-grained” patches in which the self-energy is momentum independent, but retains the full frequency dependence \cite{Maier05,Fuhrmann07}.
The method is approximate but controlled in the sense that the thermodynamic limit for local quantities are approached $\sim N_c^{-2}$ as cluster size $N_c \rightarrow \infty$ \cite{Maier05, Fuchs11, LeBlanc13, LeBlanc15}.
The eight-site cluster used here is a compromise chosen large enough to accommodate a clear nodal-antinodal differentiation \cite{Werner098site, Gull09_8site}, pseudogap regime \cite{Gull10_clustercompare}, and superconductivity \cite{Maier05_dwave, Gull13_super}, while remaining cheap enough for simulation of a wide range of parameters.

\begin{figure}[tbh]
    \centering
    \includegraphics[width=0.95\columnwidth]{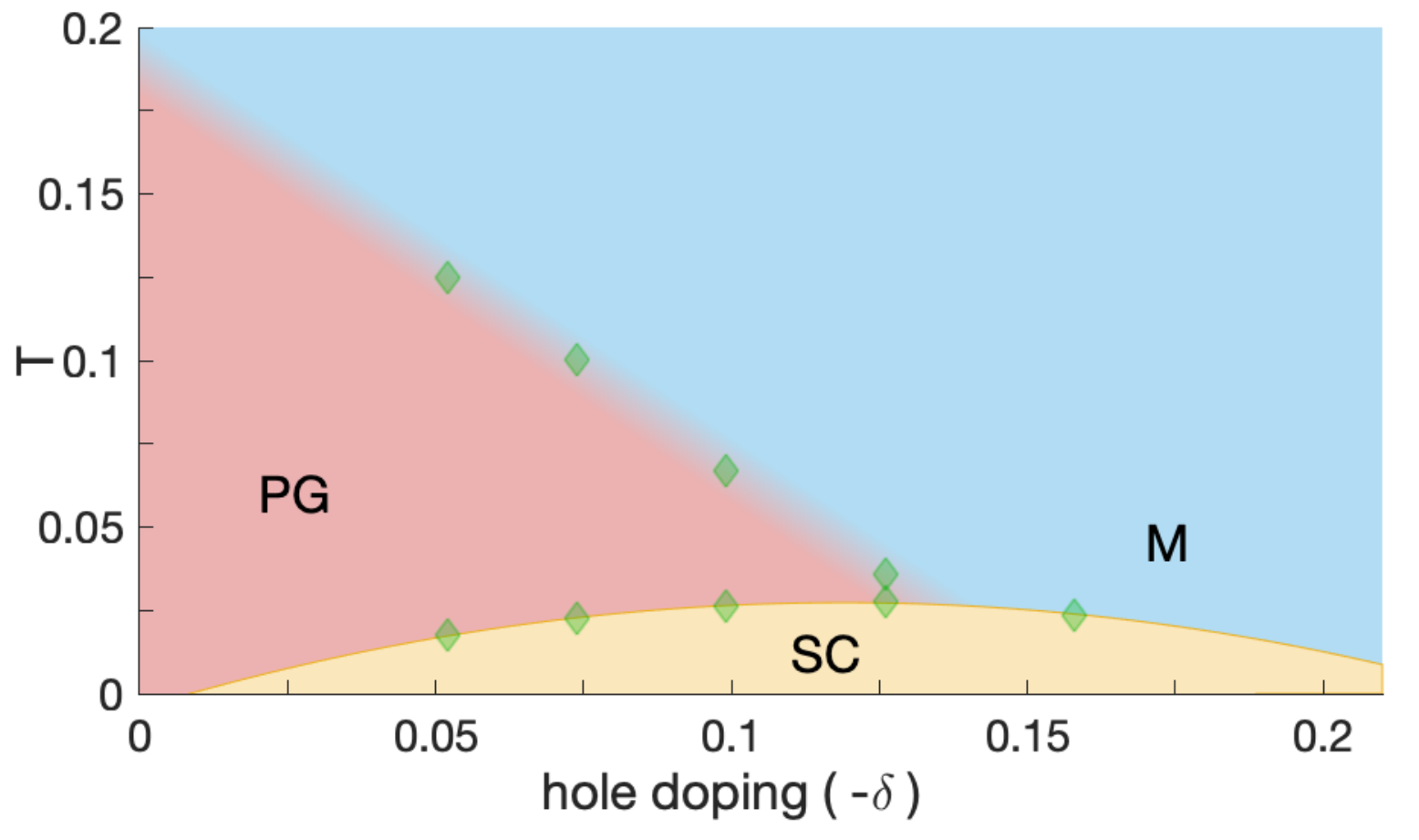}
    \llap{\put(-80,60)
     {\transparent{0.8}\includegraphics[height=2.2cm]{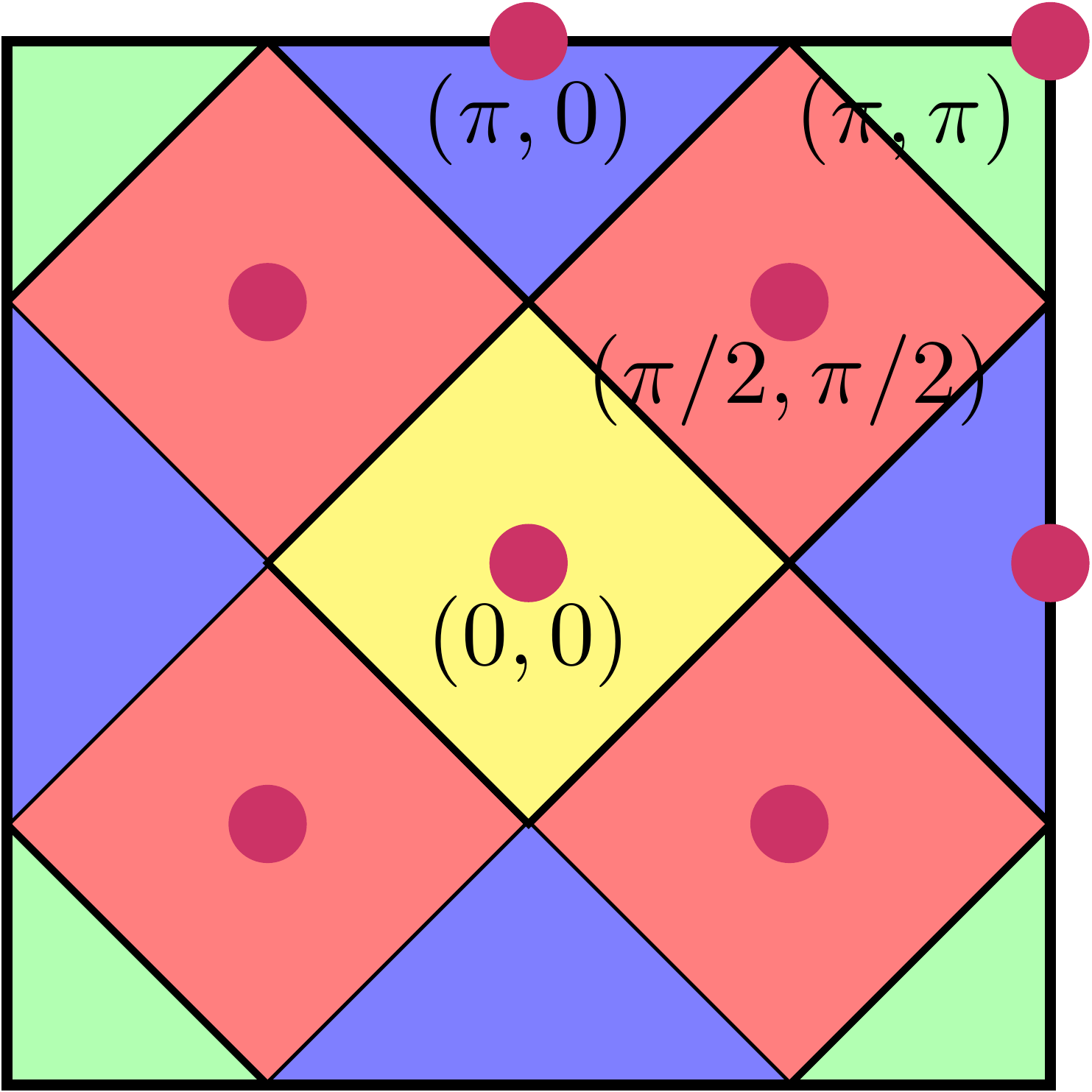}}
    }
    \caption{8-site DCA phase diagram of the Hubbard model, with metal (M; blue), pseudogap (PG; red), and superconducting (SC; yellow) regions, reproduced from Ref.~\onlinecite{Dong2019}. In the paper we use $\delta < 0$ for hole doping and $\delta > 0$ for electron doping. Inset: Geometry of the 8-site DCA cluster.}
    \label{fig:PDOverview} 
\end{figure}

Figure~\ref{fig:PDOverview} shows the phase diagram of the model obtained within the 8-site DCA with parameters $U/t=7$ and $t'=-0.15t$. 
These parameters are chosen to represent the overall phase diagram common to several cuprates \cite{Gull13_super}.
The pseudogap regime is obtained by observing a suppression in the single-particle spectral function. The superconducting phase is computed in a Nambu formulation and defined as the area where the anomalous Green's function $F(k,\tau) = -\langle T_{\tau}c_{k,\uparrow}(\tau)c_{-k,\downarrow}(0)\rangle$ at $k = (0, \pi), (\pi, 0)$ becomes nonzero.
%
%

DCA only yields the cluster Green's functions and corresponding cluster susceptibilities $\chi_{c,d}^{\omega\omega'\Omega}(K, K', Q)$ at the cluster momenta $K, K'$, and $Q$. 
The corresponding approximation to Eq.~\ref{eq:chid} is obtained by interpolating $\chi_{d}(k, k', Q) = (\chi_{0}^{-1}(k, k', Q) + \chi_{c,d}^{-1}(K, K', Q) - \chi_{c,0}^{-1}(K, K', Q))^{-1}$. This is analogous to identifying the DCA cluster vertex with the lattice vertex $\Gamma_{d}^{\omega_1\omega_2\Omega}(k_1, k_2, q)$~\cite{Maier05}.

Within the eight-site cluster, the symmetry factors we used below corresponding to $s-$, $p_x$, $d_{xy}$, and $d_{x^2-y^2}$ symmetry are defined as
$g_s(K) = 1$, 
$g_{p_x}(K) = \sin(K_x)$, 
$g_{d_{xy}}(K) $ $=$  $\sin(K_x)\sin(K_y)$, and
$g_{d_{x^2-y^2}}(K)$ $=$ $\cos(K_x)-\cos(K_y)$.
The DCA approximation generates strong antiferromagnetic (AFM) fluctuations with a correlation length comparable to the cluster size. If the establishment of long-range AFM order is allowed, the system chooses an ordered state at a temperature above the onset of the PG or superconductivity. This is a finite size effect~\cite{Mermin1966}. For this reason, we suppress magnetic long-range order and only show results obtained in the paramagnetic state, which have the correlation length of AFM fluctuations restricted to the cluster size \cite{Maier05}.

Our results are obtained with a continuous-time auxiliary field quantum Monte Carlo impurity solver \cite{Gull08,Gull11} based on the ALPS \cite{Gaenko17,Wallerberger18} libraries. The summation over fermionic frequencies in Eq.~\ref{eq:chig} and Eq.~\ref{eq:vertsymm} goes over all frequencies from $-\infty$ to $\infty$. In our calculations, only a finite number of frequencies are available, but the asymptotic behavior of $\chi_0$ is known analytically \cite{Hafermann_thesis}, . In the results presented here, we use 36 fermionic frequencies for $\beta t= 5, 10$; 50 fermionic frequencies for $\beta t= 15, 20$; and 80 fermionic frequencies for $\beta t= 30$ on both positive and negative sides to compute the vertices. The relative change for omitting the last eight frequencies on each side is on the order of $10^{-3}$. The asymptotic behavior of vertex $\Gamma$ and $F$ is also analyzed in Ref.~\onlinecite{wentzell2016highfrequency} which provides an alternative way of treating the high frequency behavior.
$\chi$ is computed with the number of frequencies listed above for the vertex correction part, plus $\chi_0$ computed with 1024 fermionic frequencies (both positive and negative) and supplemented with an analytically known asymptotic correction.

%
\begin{figure}[tbh]
    \centering
    \includegraphics[width=0.95\columnwidth]{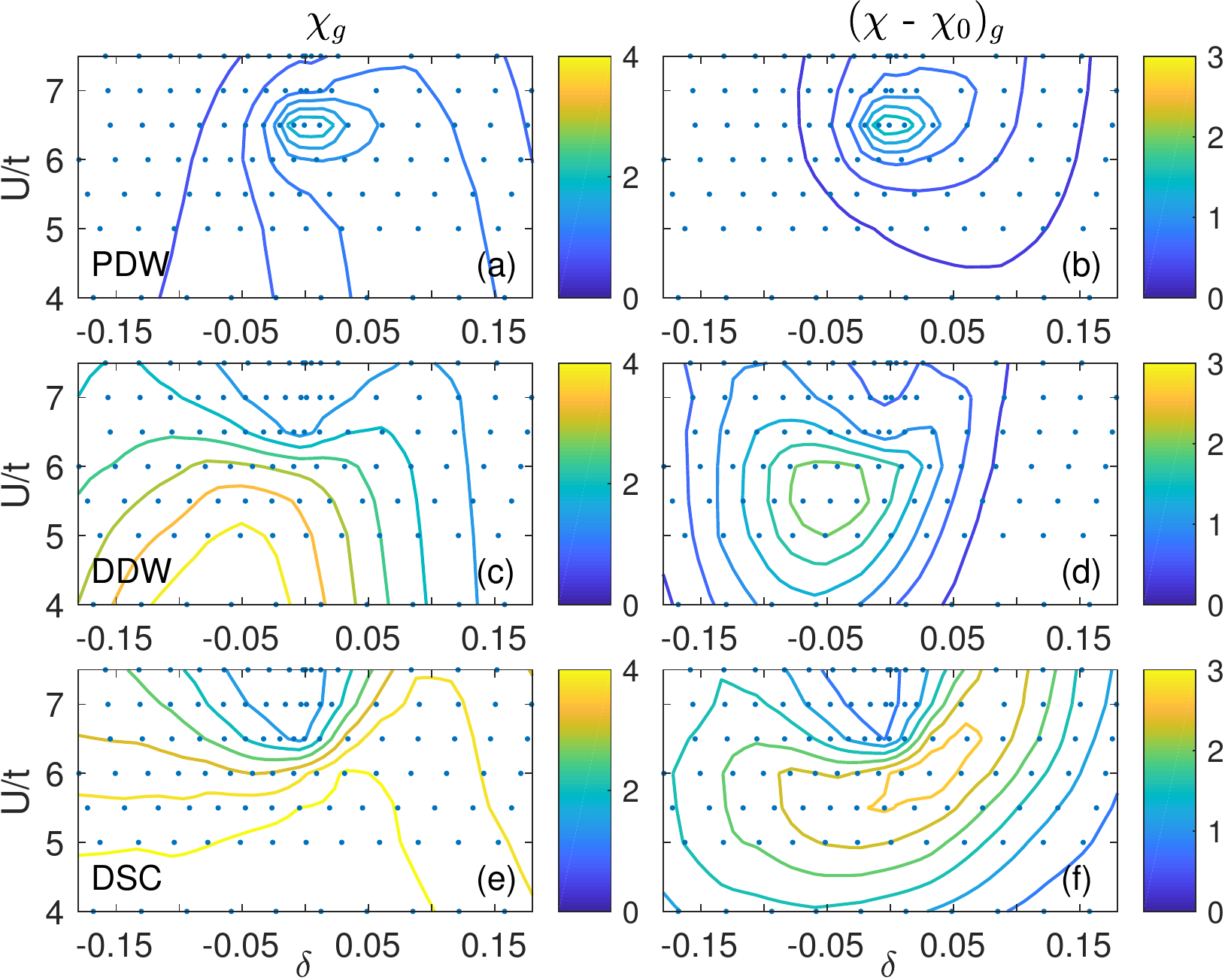}
    \caption{Amplitude of the susceptibility $\chi_g$ [left panels, Eq.~\ref{eq:chig}] and vertex corrections [right panels; Eq.~\ref{eq:vertsymm}] for charge fluctuations with $p$-wave symmetry (top), $d$-wave symmetry (middle), and $d$-wave superconductivity (bottom) as a function of doping $\delta$ and interaction strength $U$ at $\beta t = 20$, $t' = -0.15~t$. Color bars show the strength of the susceptibility.}
    \label{fig:chivert20} 
\end{figure}
%

\section{Charge and Superconducting fluctuations} \label{sect: fluct}
Figure. \ref{fig:chivert20} shows six panels for the leading fluctuations at $\beta t = 20$ and $t'/t = -0.15$. The three rows represent $p_x$ and $d_{x^2 - y^2}$ density fluctuations, and $d_{x^2 - y^2}$ superconducting fluctuations. The left columns show $|\chi_g|$ (see Eq.~\ref{eq:chig}), the right columns  $|(\chi - \chi_0)_g|$ (see Eq.~\ref{eq:vertsymm}). Each panel displays data as a function of $U$ and doping $\delta$, with $\delta = 0$ corresponding to half filling. For superconductivity, only fluctuations with $d_{x^2 - y^2}$ symmetry (DSC) are large \cite{Chen2015}. 
%
Fig.~\ref{fig:chivert20} (b), (d) and (f) illustrate that $\chi$ itself is not a good measure for the correlation contribution that may eventually drive the system to an ordered state, as most of $\chi$ stems from $\chi_0 = G G$.
Fig. \ref{fig:chivert20} (c)-(f) show that the amplitudes of DDW and DSC fluctuations are comparable (at the same order), implying competing fluctuations. However, we find numerically that DSC fluctuations are always larger than DDW for the parameters examined. 
In a small regime of parameter space, where $U \sim 6.5t$ with slight electron doping, p-density wave fluctuations are the dominant charge fluctuation (panel (a) and (b)). 

The maximum of DSC is on the electron-doped side, while the maximum of DDW is on the hole-doped side, both at intermediate interaction strength $U$. In addition, DSC fluctuations are suppressed in the psedogap regime starting from $U \sim 6t$ (see Refs. \onlinecite{Chen2015, Gull09_8site}), while DDW fluctuation starts to show suppression around half filling for $U \sim 6.5t$, which corresponds to the onset of the Mott insulator \cite{Gull09_8site} in this approximation. PDW does not show any suppression by either the pseudogap or the Mott insulating state; its maximum is near $U \sim 6.5t$, which is the same interaction strength where DDW shows a suppression near half filling.

\section{Temperature evolution} \label{sect: temp}
\begin{figure}[tbh]
    \centering
    \includegraphics[width=0.95\columnwidth]{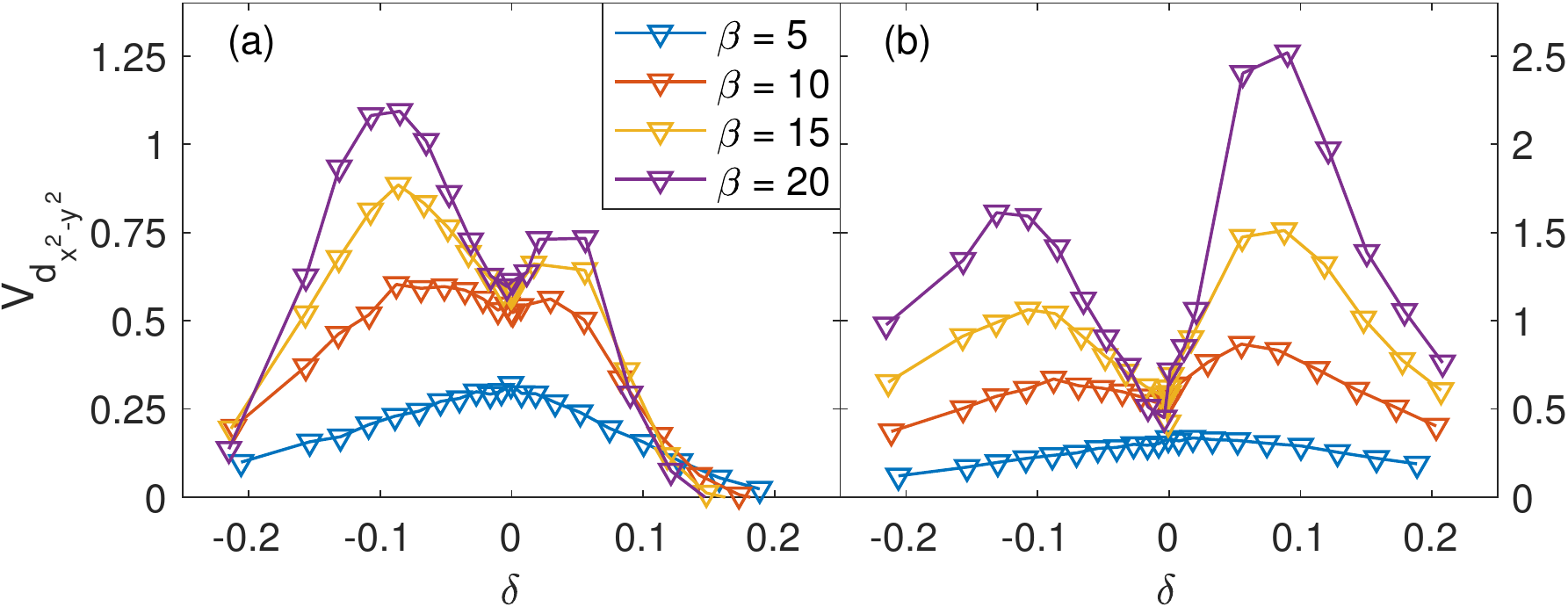}
    \caption{Amplitude of vertex correction for DDW and DSC at $U = 7~t$, $t' = -0.15~t$ for 4 temperatures. We use $t = 1$ here. (a): DDW. (b): DSC. Note that the y ranges of panels (a) and (b) are different.}
    \label{fig:vert_temp} 
\end{figure}

In order to investigate the competition between $d$ density wave and $d$ wave superconducting fluctuations in more detail, we explore their temperature evolution with different dopings in Fig.~\ref{fig:vert_temp}. 
We show the results at $U = 7~t$ and $t' = -0.15t$.
Panel (b) shows that away from half filling, the vertex part of DSC increases as temperature decreases for all doping levels investigated (see also Ref.~\onlinecite{Chen2015}), whereas panel (a) shows that the vertex correction part of DDW increases as temperature decreases in the underdoped regime away from half filling but rapidly decays to zero for large doping. The maximum of the DSC fluctuations are near the maximum $T_c$ \cite{Chen2015}, while the corresponding maximum DDW fluctuations occur at slightly lower doping. The transition to superconductivity on the hole-doped side will take place near $\beta t = 35$ at optimal doping in this model, {\it i.e.}, at a temperature about twice below where these results have been obtained.

\begin{figure}[tbh]
    \centering
    \includegraphics[width=0.95\columnwidth]{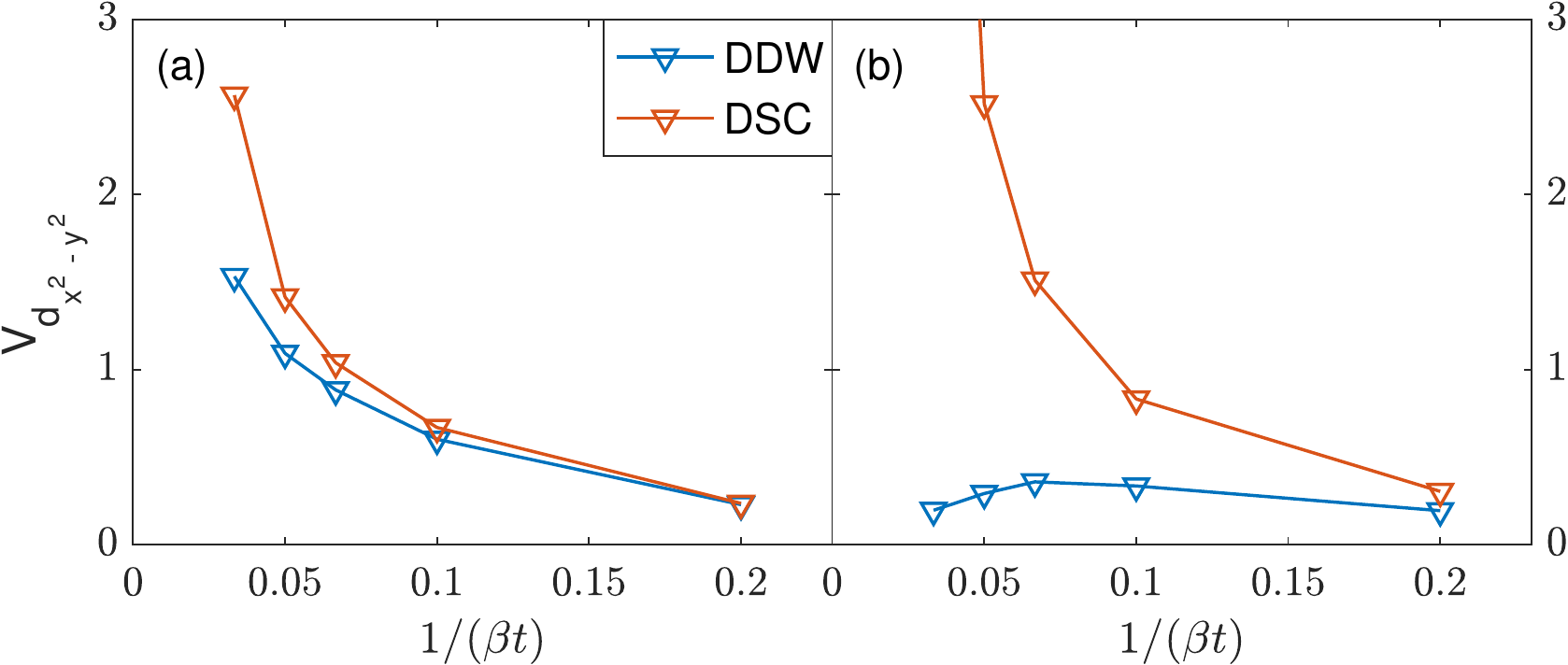}
    \caption{Temperature evolution of amplitude of vertex correction for DDW and DSC at $U = 7~t$, $t' = -0.15~t$, (a) $\mu/t=-1.4$ corresponding to $\delta \sim -0.09$ and (b)  $\mu/t=1$ corresponding to $\delta \sim 0.09$.}
    \label{fig:vert_temp_extra} 
\end{figure}

The amplitude of the vertex correction part of DDW and DSC fluctuations as a function of temperature at the doping level corresponding to largest DDW fluctuation and largest DSC fluctuation is shown in Fig.~\ref{fig:vert_temp_extra}. These results are obtained in the paramagnetic state but reach temperatures just above the superconducting transition. At the doping level where DDW fluctuations are strongest (corresponding to $\mu/t=-1.4, \delta \sim 0.09$), DSC fluctuations are substantially larger than DDW, and increase faster as temperature decreases. We have been unable to find a region of parameter space where $d$ density wave order prevails over superconductivity around optimal doping. At the doping level corresponding to the largest DSC fluctuation($\mu/t=1, \delta \sim 0.09$) we could find, DDW fluctuations first increase as temperature decreases, then start to decrease at $\beta t\sim 20$. This result is consistent with the findings of Refs.~\onlinecite{Macridin2004, Honerkamp2002}.

%
%

%
\begin{figure}[tbh]
    \centering
    \includegraphics[width=0.95\columnwidth]{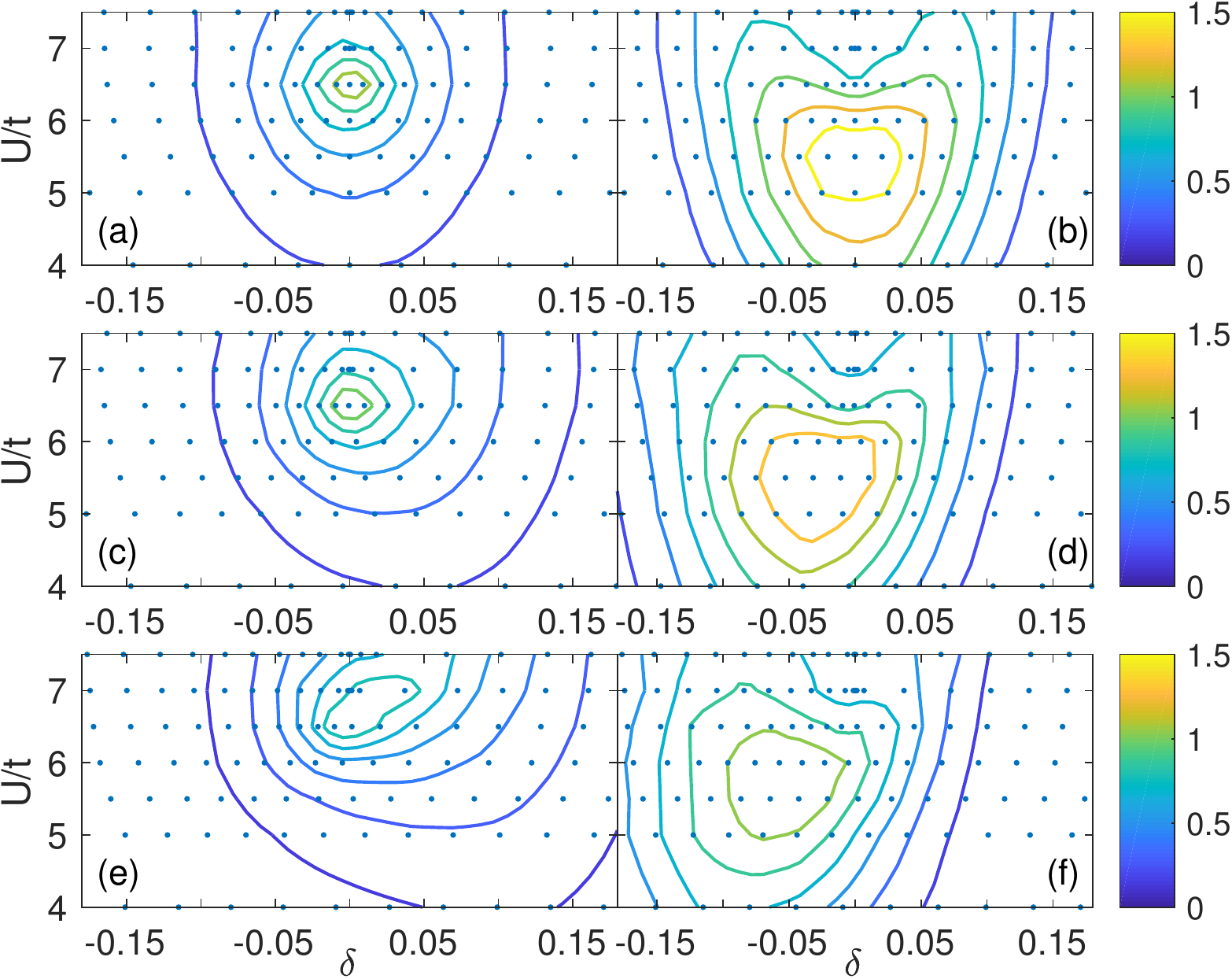}
    \caption{Amplitude of vertex correction for $p$ density wave and $d$ density wave at $\beta t= 15$ and different $t'$. (a): $p$ density wave, $t'/t = 0$. (b): $d$ density wave, $t'/t = 0$. (c): $p$ density wave, $t'/t = -0.10$. (d): $d$ density wave, $t'/t = -0.10$. (e): $p$ density wave, $t'/t = -0.20$. (f): $d$ density wave, $t'/t = -0.20$. The small lack of reflection symmetry of the results at $t' = 0$ is due to Monte Carlo errors.}
    \label{fig:vert_tprime} 
\end{figure}

%
\begin{figure}[tbh]
    \centering
    \includegraphics[width=0.95\columnwidth]{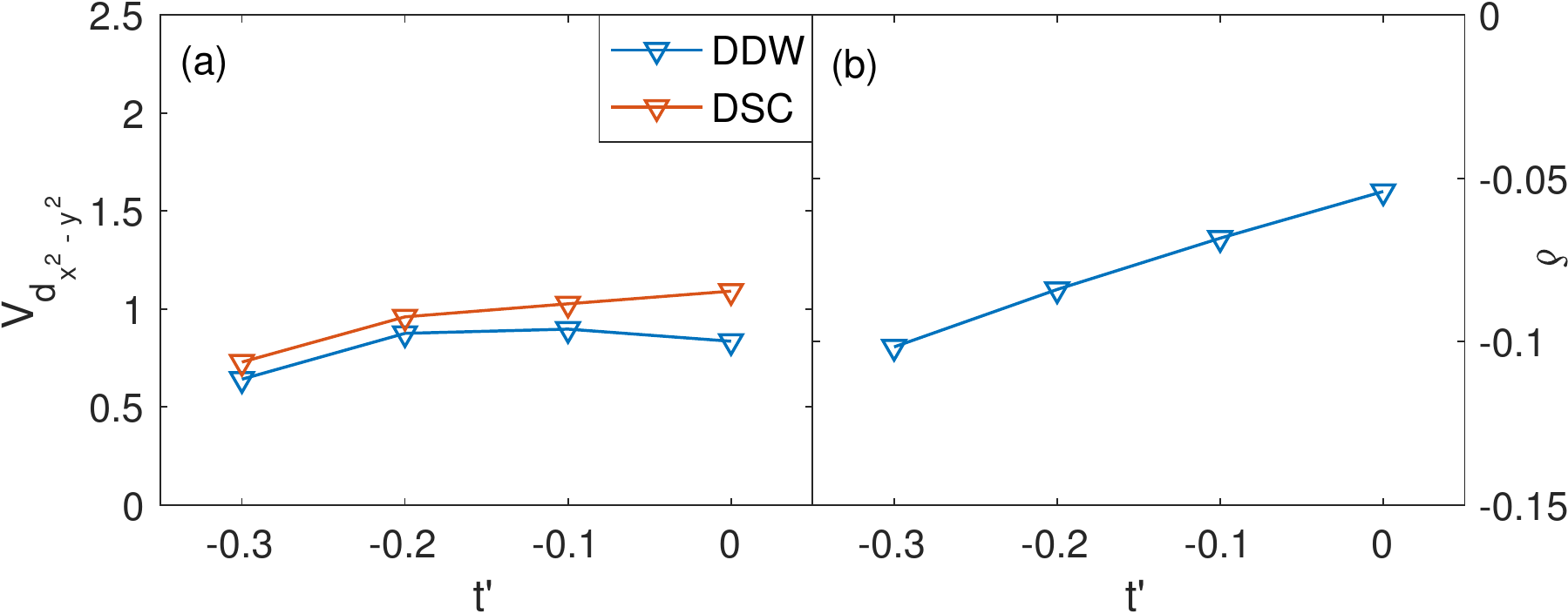}
    \caption{(a) $t'$ evolution of vertex correction with $d_{x^2 - y^2}$  symmetry [$V_{d_{x^2 - y^2}}$, Eq.~\ref{eq:vertsymm}] for DDW and DSC at $U = 7~t$, $\beta t = 15$ at doping levels corresponding to largest DDW fluctuations. (b) Doping levels corresponding to largest DDW fluctuations.}
    \label{fig:vert_tprime_extra} 
\end{figure}

\section{Particle-hole asymmetry} \label{sect: ph}
Figure.~\ref{fig:vert_tprime} shows the dependence of PDW and DDW on interaction and doping for three values of $t'$. $t'$ is to shift the Van Hove singularity of the density of states toward hole doping and destroys particle hole symmetry.
We find that as $t' / t$ is changed from $0$ to $-0.2$ and correspondingly the Van Hove singularity more toward the hole-doped side, PDW fluctuations spread out over a larger area of the parameter space at the electron-doped side, but their maximum does not change significantly.
On the other hand, the maximum of DDW fluctuations shifts toward hole doping and their intensity decreases substantially.
The particle-hole asymmetry of DSC shows a different trend from DDW \cite{Chen2015}. As $ -t' / t$ increases, the maximum of DSC fluctuations moves to the electron-doped side. This is consistent with a scenario where the  establishment of a pseudogap on the hole-doped side suppresses $d$-wave superconducting fluctuations \cite{Chen2015}. 

Panel (a) of Fig.~\ref{fig:vert_tprime_extra} shows the amplitude of the vertex corrections of DDW and DSC with different $t'$ at $U = 7t$, $\beta t= 15$, and doping levels corresponding to the largest DDW fluctuations. As $-t'$ increases, the maximum of DDW shifts toward the hole-doped side. Since the overall intensity of the fluctuations decreases, the maximum value of DDW fluctuations saturates and then drops. At the same time, for all the $t'$ values we explore, DDW fluctuations are still weaker than DSC fluctuations around the optimal doping for DDW fluctuations.
Panel (b) shows the doping levels where the largest DDW fluctuations are found.

Eight-site DCA shows a narrowing of the pseudogap on the electron-doped side, and for $t' < -0.15t$ has a first-order transition between the Mott insulator and a momentum-dependent Fermi liquid state \cite{Gull09_8site}. This rapid change in the single-particle quantities does not show an analog in the DDW or PDW fluctuations.

\section{Discussion} \label{sect: diss}
The causes and consequences of DDW have been much debated. Early studies \cite{Chakravarty2001, Honerkamp2001, Tewari2001, Morr2002, Morr2003} considered such fluctuations as a candidate for the mechanism of the pseudogap. Numerical calculations of the Hubbard model \cite{Stanescu2001, Honerkamp2002, Macridin2004, Lu2012, Otsuku2014} argued that this type of fluctuation is not strong enough to form an ordered state in the strongly correlated region, and is always dominated by DSC fluctuations, while renormalization group studies find that for SU(N), when $N > 6$, DDW becomes the leading instability \cite{Honerkamp2004}.
Other works have studied the coexistence of DDW and DSC orders \cite{Tewari2001, Ismer2006}.
The $d$-wave fluctuations at $Q = (0, 0)$ or $d$-nematic fluctuations have  been studied in Ref~.\onlinecite{Honerkamp2002}, where it is shown that they grow together with DDW, DSC, and antiferromagnetic fluctuations.

More recent variational Monte Carlo studies \cite{Yokoyama2016, Kobayashi2017} found staggered flux states (DDW states) in the strongly correlated underdoped regime of the Hubbard model, for interactions smaller than the Mott transition. They proposed this state as a candidate for an anomalous ``normal state'' competing with DSC since its properties are similar to those of the pseudogap. In these more recent works, the state does not coexist with DSC.

Our results clarify some of these arguments. 
Our interaction, doping, and $t'$ evolution of DDW correlations show that the evolution of the pseudogap and that of DDW correlations do not track each other. 
This can be seen from the fact that in the area around half filling, where there is a pseudogap, DDW correlations show a suppression. However, the interaction strength where DDW starts to show a suppression is at $U \sim 6.5t$, not at $U \sim 6t$, where the pseudogap opens.
As $-t'/t$ increases, the pseudogap regime moves to the hole-doped side (shown in Fig.~3 and Fig.~4 of Ref.~\onlinecite{Gull09_8site}), while the suppression of DDW moves to the electron doped side. These trends show little overall correlation between areas with the largest DDW fluctuations and the appearance of a pseudogap. 

Without entering an ordered phase, it is difficult to make statements about a potential coexistence of DDW and DSC. However, comparison between the temperature evolution of DDW and DSC at $U = 7t$ (Fig.~\ref{fig:vert_temp} and Fig.~\ref{fig:vert_temp_extra}) clearly shows that around the optimal doping for DDW, DSC correlations are dominant over DDW and will order first. Thus, if there is a coexistence regime, it is likely fully contained inside the DSC dome.

Previous works \cite{Macridin2004} investigated the competition between DSC and DDW in the Hubbard model with DCA on 4-site clusters at the two doping levels $\delta = -0.05$ and $\delta = -0.25$. The first doping level is in the pseudogap regime, the second far in the overdoped.
The main finding, namely that the susceptibility corresponding to the $d$ density wave does not diverge, indicating the absence of a possible transition to the DDW state, is reproduced by our calculations. However, the claim that both DDW and DSC correlation functions are enhanced in the pseudogap regime is inconsistent with our more detailed calculations.
The reason is that the intermediate maximum, which we find around $\delta = -0.09$, is missed by the coarse doping resolution employed in Ref.~\onlinecite{Macridin2004}.

DDW, the staggered flux (SF) state, and its competition with DSC, have been studied extensively in the t-J model \cite{Ubbens1992, Lee1998, Cappelluti1999, Leung2000, Ivanov2000, Ivanov2003, Hamada2003, Bejas2011, Bejas2012}, which maps to the Hubbard model in certain limits. In the t-J model, a well-defined SF state exists in a small doping region in the phase diagram \cite{Bejas2011, Bejas2012}.
Large-N calculations on the model~\cite{Bejas2011} related the DDW self-energy to the pseudogap via the formation of Fermi arcs. This is different from the results presented here, but whether the differences should be attributed to the different models, parameter regimes, approximations, or post-processing procedures is a question for future study. 

Fluctuation diagnostics \cite{Gunnarsson15} attributes contributions to the self-energy to fluctuations of different types. Reference.~\onlinecite{Dong2019} shows that in the parameter regime we explore in this paper, the dominant fluctuation in the normal state is always magnetic. It is shown in Ref.~\onlinecite{Gunnarsson15} that these magnetic fluctuations are responsible for the major contribution to the single-particle self-energy, and thereby for the suppression of the density of states. The pseudogap in the Hubbard model can therefore clearly be attributed to magnetic fluctuations. 
DDW and DSC fluctuations do not directly contribute to a single-particle self-energy, since the summation of a fluctuation with a $d$-wave symmetry tends to cancel out in the calculation of the self-energy \cite{Gunnarsson15}.

The DCA simulations performed here are insensitive to the stripe order widely found in experiment and in numerical ground-state calculations. In order to find a transition to an ordered state in DCA, the ordering vector typically needs to be commensurate with the cluster geometry. However, the stripe orders, {\it e.g.}, found in Ref.~\onlinecite{Zheng17_stripe} are too large to fit into the DCA cluster studied here. Thus, while the method is sensitive to charge fluctuations on a length scale smaller than the cluster size, DCA is not expected to find the period 4 and period 5 stripes of Ref.~\onlinecite{Zheng17_stripe}. Appropriately chosen larger clusters may find these stripes, but finite size effects would likely overestimate their contribution. The unbiased detection of such orders with DCA or lattice methods is an important open problem.
Our conclusion on the relation between PG and DDW, however, is not influenced by the omission of the stripe order. With an eight-site cluster simulation, we can see a clear signal of a pseudogap from the single particle spectrum, while DDW fluctuations have been shown to be unrelated to it. 


\section{Conclusions} \label{sect: conclu}
In conclusion, we have investigated the physics of short-range charge fluctuations in the strongly correlated regime of the 2D Hubbard model within the 8-site DCA approximation.
We have found that the dominant charge fluctuations have $d$-wave symmetry, apart from a small regime near the Mott transition, where we find $p$-wave charge fluctuation.
For all doping, interaction, and next-nearest-neighbor hopping parameters investigated, we showed that superconducting $d$-wave fluctuations are always stronger than charge fluctuations away from half filling. At $U = 7t$, at the doping level that is most favorable for DDW, DSC will order first as temperature decreases, showing that if there were a coexistence of DDW and DSC orders, the coexistence area would likely be fully contained inside the DSC dome.

Our parameter scans show that DDW fluctuations cannot be viewed as the cause of the pseudogap in the single-particle density of states, as the change of DDW fluctuations does not match the evolution of the pseudogap. This is consistent with the results of several recent works, including Refs.~\onlinecite{Gunnarsson15,Dong2019}, that show convincingly that the pseudogap can be attributed to strong short-wavelength AFM correlations.


\acknowledgments{
This work was supported by NSF DMR-1606348 and the Simons Collaboration on the Many-Electron Problem. We thank Kai Sun for insightful discussions.
}
\bibliographystyle{apsrev4-1}
\bibliography{refs}
\end{document}